\documentclass[12pt]{article}
\usepackage{epsf}
\usepackage{epsfig}
\usepackage{wrapfig}
\usepackage{amsmath}
\usepackage{amssymb}
\usepackage{eufrak}
\usepackage{graphicx}
\begin{document}
\begin{center}
\section*{Fidelity, quantum computations and Wilson loop}
\vskip 5mm V. I. Kuvshinov$^{1}$, P. V. Buividovich$^{2}$ \vskip
5mm {\small (1) {\it  JIPNR, National Academy of Science, Belarus,
220109 Minsk, Acad. Krasin str. 99}
\\
(2) {\it Belarusian State University, Minsk, F. Skoriny av. 4} }
\end{center}
\vskip 5mm
\begin{abstract}
General properties of quantum systems which interact with
stochastic environment are studied with a strong emphasis on the
role of physical symmetries. The similarity between the fidelity
which is used to characterize the stability of such a systems and
the Wilson loop in QCD is demonstrated, and the fidelity decay
rates are derived. The consequences of existence of the symmetry
group on the statistical properties of the system are analyzed for
various physical systems - a simple quantum mechanical system,
holonomic quantum computer and Yang-Mills fields.
\end{abstract}
\vskip 8mm
\subsection*{Introduction}

\indent \indent Transition between classical and quantum behaviour
is an important phenomenon in different fields of physics.
Sometimes this transition is associated with the processes of
decoherence which are commonly described as interaction between
quantum system and environment in the state of thermodynamical
equilibrium \cite{Haake, Liouville, Gardiner}. It is known that
the environment can be described as the stochastic ensemble of
classical fields \cite{Sim1, Haake, Prosen, Liouville, Gardiner}.
Therefore it is important to analyze the behavior of quantum
systems which interact with random fields and to relate the
statistical properties of random fields to the physical properties
of the environment, for example, the properties of symmetry.
Implementations of the model of stochastic environment with
definite symmetry properties are very diverse - for example, the
stochastic vacuum of Yang-Mills field which is known to possess
confining properties, or the errors in quantum computations. The
stability of quantum motion is usually characterized in terms of
the fidelity. In this paper the expression for the fidelity of
holonomic quantum computations is derived, and the analogy between
the theory of holonomic quantum computations and the theory of
gauge fields is demonstrated. It is shown that the roots of this
analogy lie in the existence of invariance group of the system.
This analogy is used to introduce the concept of the fidelity in
QCD. It is shown that the fidelity decay rate essentially depends
on the "string tension". Invariance of the stochastic ensemble of
classical fields under the action of the symmetry group is used to
simplify the calculations.
\subsection*{1. Multiplicative integrals with random variables}
\indent \indent The quantities which are analyzed in this paper
are most naturally expressed in terms of the expectation values of
multiplicative integrals over some subspaces in the configuration
space of the problem. Multiplicative integral of operator field
$\hat{A}\left({ \bf x} \right)$ over some manifold $D$ of
dimensionality $d$ is defined as \cite{Gantmacher}:
\begin{equation} \label{MultiplInt} \hat{U}\left[
\hat{A}\left({\bf x} \right) \right] = \hat{P} \exp \left ( i \int
\limits_{D} dV \hat{A}\left({\bf x} \right)\right) =
\lim_{N\rightarrow\infty}\prod^{N}_{k=1} \left(\hat{I} + i \hat{A}
\left( {\bf x}_{k}\right) \Delta V_{k} \right)
\end{equation}
where $\hat{P}$ is the path-ordering operator, $\hat{I}$ is the
identity operator, $dV$ is the volume element on $D$ and $\Delta
V_{k}$ is the small finite element of volume on $D$. Both $dV$ and
$\hat{A}\left({\bf x} \right)$ can be tensors, but their product
should be a scalar quantity.
\\ \indent Generally there is no
expression in closed form for the expectation value of
multiplicative integral of an arbitrary random field
$\hat{A}\left({\bf x} \right)$, but a good approximation can be
found when the size of the manifold $D$ considerably exceeds the
correlation length $l_{\rm corr}$ of random field. In this case
the manifold $D$ can be split into a large number $m$ of
submanifolds $D_{k}$, $k=1 \ldots m$ with the sizes still much
greater than the correlation length. The multiplicative integral
(\ref{MultiplInt}) is then rewritten as the product of
multiplicative integrals over each submanifold $D_{k}$. As the
sizes of submanifolds $D_{k}$ exceed the correlation length of the
random field, multiplicative integrals over them can be treated as
statistically independent variables, and expectation value of
their product can be calculated as the product of expectation
values of each factor. Each multiplicative integral over
submanifold $D_{k}$ is then expanded in powers of
$\hat{A}\left({\bf x} \right)$. The result of these
transformations is:
\begin{equation}\label{MultiplIntExpansion}\begin{array}{ll}
\overline{\hat{U}} =
\prod\limits_{k=1}^{m}\hat{P}\overline{\exp\left(i\int\limits_{D_{k}}dV\hat{A}\left({\bf x}\right)\right)}=\\
\prod\limits_{k=1}^{m} \left(\hat{I} + i \int \limits_{D_{k}}dV
\overline{\hat{A}\left({\bf x}\right)} - 1/2 \: \hat{P}\int
\limits_{D_{k}}\int \limits_{D_{k}}dV_{1}dV_{2}
\overline{\hat{A}\left({\bf x}_{1}\right) \hat{A}\left({\bf
x}_{2}\right)} + \ldots \right)
\end{array}\end{equation}where we have used the fact that the operation of path-ordering is
linear and can be therefore exchanged with averaging. The function
$\hat{C} \left({\bf x}, {\bf y} \right)=
\overline{\hat{A}\left({\bf x}\right)\hat{A}\left({\bf y}\right)}
-\overline{\hat{A}}\left({\bf
x}\right)\overline{\hat{A}}\left({\bf y}\right)$ is the
second-order correlation function of random field
$\hat{A}\left({\bf x}\right)$. Here we will consider only
statistically homogeneous random fields. A random field is
statistically homogeneous if its correlation functions depend only
on the differences of their arguments \cite{Stat}. In this paper
the following form of the correlation function is assumed:
\begin{equation}
\label{CorrelationFunction} \hat{C}\left({\bf x},{\bf
y}\right)=\hat{C}f\left({\bf x} - {\bf y}\right), \quad
f\left({\bf 0}\right)=1
\end{equation}
The function $f\left({\bf x}\right)$ is significantly large only
when $|{\bf x}|<l_{\rm corr}$. The second-order term in the
expansion (\ref{MultiplIntExpansion}) is then approximated to the
first order in the volume $\Delta V_{k} = \int \limits_{D_{k}} dV$
of submanifold $D_{k}$ as $\hat{P}\int \limits_{D_{k}}\int
\limits_{D_{k}}dV_{1}dV_{2} \overline{\hat{A}\left({\bf
x}_{1}\right) \hat{A}\left({\bf x}_{2}\right)} \approx 1/2 \:
l_{\rm corr}^{d}\hat{C}\Delta V_{k}$. High-order terms in this
approximation are estimated as $\hat{P}\int\limits_{D_{k}} \ldots
\int\limits_{D_{k}} dV_{1} \ldots dV_{i}
\overline{\hat{A}\left({\bf x}_{1}\right) \ldots \hat{A}\left({\bf
x}_{i}\right)} \sim \hat{C}^{\: i/2}l_{\rm
corr}^{d\left(i-1\right)}\Delta V_{k}$. High-order terms can be
neglected if the following condition holds:
\begin{equation}
\label{GaussianDominance} \hat{C}^{\: 1/2}l_{\rm corr}^{d} \ll 1
\end{equation}
This condition is called the condition of gaussian dominance
\cite{Sim1,Sim2}. For the sake of simplicity we will assume that
this condition holds. The contribution of high-order terms can be
taken into account without any difficulties. Under these
assumptions the expression (\ref{MultiplIntExpansion}) can be
rewritten in a compact form using the definition of multiplicative
integral (\ref{MultiplInt}) and the fact that the number $m$ is
large:
\begin{equation}
\label{AveragedMultiplIntFinal} \overline{\hat{U}} \cong
\prod\limits_{k=1}^{m} \left(\hat{I} + i
\overline{\hat{A}\left({\bf x}_{k} \right)}\Delta V_{k} - 1/2 \:
l_{\rm corr}^{d}\hat{C}\Delta V_{k}\right) \cong \hat{P}\exp\left(
\int \limits_{D} dV \left( i \overline{\hat{A}\left({\bf
x}\right)} - 1/2 \: l_{\rm corr}^{d}\hat{C} \right)\right)
\end{equation}
\indent If the stochastic ensemble of fields $\hat{A}\left({\bf x}
\right)$ is invariant under global transformations $\hat{T}$ which
build an irreducible representation of some group $G$, the
expectation value of the multiplicative integral
(\ref{MultiplInt}) is proportional to the identity. Invariance of
the stochastic ensemble is understood as the following
identity:\begin{equation} \label{InvarianceCondition} DP\left[
\hat{A}\left({\bf x} \right) \right]=DP \left[ \hat{T} \hat{A}
\left( {\bf x} \right) \hat{T}^{-1} \right]
\end{equation}where  $DP\left[ \hat{A}\left({\bf x} \right) \right]$ is the
probability measure on the stochastic ensemble of fields
$\hat{A}\left({\bf x} \right)$. This assertion is proved by
applying an arbitrary transformation from the group $G$ to the
expectation value of the product of an arbitrary number of field
variables $\hat{A}\left({\bf x} \right)$:
\begin{equation}
\label{InvarianceProof}
\begin{array}{ll}
\hat{T}^{-1}\overline{\hat{A}\left( {\bf x}_{1}\right) \ldots
\hat{A}\left( {\bf x}_{i}\right) }\hat{T} = \int DP \left[ \hat{A}
\left({\bf x} \right) \right] \left(\hat{T}^{-1}\hat{A}\left( {\bf
x}_{1}\right)\hat{T} \ldots
\hat{T}^{-1}\hat{A}\left( {\bf x}_{i}\right)\hat{T} \right)= \\
\int DP \left[\hat{T}^{-1} \hat{A} \left({\bf x} \right) \hat{T}
\right] \left(\hat{T}^{-1}\hat{A}\left( {\bf x}_{1}\right)\hat{T}
\ldots \hat{T}^{-1}\hat{A}\left( {\bf x}_{i}\right)\hat{T} \right)
= \overline{\hat{A}\left( {\bf x}_{1}\right) \ldots \hat{A}\left(
{\bf x}_{i}\right) }
\end{array}
\end{equation}
Thus the expectation value of the product of an arbitrary number
of field variables commutes with an arbitrary transformation from
the irreducible representation of the group $G$. But then the
expectation value of the multiplicative integral
(\ref{MultiplInt}) also commutes with all such transformations.
This is simply proved by expanding the multiplicative integral in
powers of its argument and applying the transformation to each
summand. Schur's lemma in group theory states that if the linear
operator commutes with all transformations from the irreducible
representation of the group it is proportional to the identity
\cite{Elliot}. The assertion is proved. It is evident that in this
case path-ordering is no longer necessary in
(\ref{AveragedMultiplIntFinal}).
\subsection*{2. The fidelity. A simple example: quantum system with random hamiltonian}
\indent \indent The fidelity of an arbitrary quantum system is
defined as the scalar product of the state vectors which
correspond to perturbed and unperturbed states of the system. If
there are random variables in the problem, the expectation value
of the square of the absolute value of the fidelity is considered
\cite{Kuvshinov, Prosen, Liouville}:
\begin{equation}
\label{FidelityDefinition} f_{1} = \langle \phi_{0}
|\phi_{1}\rangle, \quad \quad f_{2} = \overline{|f_{1}|^{2}} =
{\rm Tr} \: \left(\hat{\rho}^{1}\hat{\rho}^{0} \right)
\end{equation}
where $| \phi_{1} \rangle$, $| \phi_{0} \rangle $ are the state
vectors of perturbed and unperturbed systems and $\hat{\rho}^{1}$
and $\hat{\rho}^{0}$ are the density matrices of perturbed and
unperturbed systems. The fidelity is used to characterize the
stability of quantum systems with respect to perturbations
\cite{Kuvshinov, Prosen, Liouville}.
\\ \indent In this section a simple example
of quantum system which interacts with stochastic process with
definite symmetry properties is considered. On this simple example
we will demonstrate that it is possible to use the expectation
value of the fidelity $f_{1}$ to estimate the fidelity decay rate
and that decay rates of the expectation value of the fidelity
$f_{1}$ and the fidelity $f_{2}$ are equal up to some coefficient
which is close to one.
\\ \indent Statistical averaging in quantum mechanics
is performed on the density matrix of the system. Evolution of the
density matrix is described by the Liouville equation ($\hbar = 1
$):
\begin{equation}
\label{LiouvilleEquation} \partial_{t} \rho _{ij} = i \left( {\rho
_{ik}H_{kj} - H_{ik}\rho _{kj}} \right)
\end{equation}
We will suppose that the hamiltonian $H_{ij}$ is the sum of the
unperturbed stationary hamiltonian and the random perturbation,
which is the stationary stochastic process \cite{Liouville,
Gardiner}. The total number of independent state vectors of the
system is assumed to be $N$. We will also suppose that the
matrices of the perturbation part of the hamiltonian form the
gaussian orthogonal ensemble \cite{RMT} which is invariant under
the action of the orthogonal group $O(N)$:
\begin{equation}
\label{StochasticHamiltonian}
\begin{array}{ll}
H_{ij} = H_{ij}^{0} + \delta H_{ij}, \quad H_{ij}^{0} = {\rm diag}
\: \left(E_{i} \right), \\ \overline{\delta H_{ij}\left(t_{1}
\right)\delta H_{kl}\left(t_{2} \right)} = \sigma ^{2}\left(
{\delta _{ik} \delta _{jl} + \delta _{il} \delta _{jk}}
\right)f\left(t_{2} - t_{1}\right), \quad \int
\limits_{0}^{\infty}dt f\left(t\right)=\tau_{\rm corr}
\end{array}
\end{equation}
where $\tau_{\rm corr}$ is the correlation time of the
perturbation. Solution of the Liouville equation
(\ref{LiouvilleEquation}) can be represented in terms of
multiplicative integral after introducing the linear operator
$X_{ijkl} = H_{lj}\delta _{ik} - H_{ik} \delta _{lj}$ on the space
of $\left[ N\times N \right]$ matrices. The Liouville equation
(\ref{LiouvilleEquation}) is rewritten as $\partial_{t} \rho _{ij}
= i X_{ijkl}\rho _{kl}$, and its formal solution is obtained by
applying the multiplicative integral of the operator $X_{ijkl}$ to
the density matrix of the initial state:
\begin{equation}
\label{LiouvilleSolution} \rho\left(t \right) = \hat{P} \exp
\left(i \int \limits_{0}^{t} d\xi \hat{X}\left(\xi \right) \right)
\rho \left(0 \right)
\end{equation}
Now it is possible to use the results of section 1 and to
calculate the density matrix which is averaged over all
implementations of stochastic process $\delta H_{ij}$. The
manifold $D$ is the time interval from $0$ to $t$, and the
condition of gaussian dominance (\ref{GaussianDominance}) is
$\sigma \tau_{\rm corr} \ll 1$. The expectation value of the
operator $X_{ijkl}$ is $\overline{X}_{ijkl} = H_{lj}^{0}\delta
_{ik} - H_{ik}^{0} \delta _{lj}$, the second-order correlation
function is $C_{ijmn} = \overline{X_{ijkl}X}_{klmn} -
\overline{X}_{ijkl}\overline{X}_{klmn} = 2 \sigma^{2} \left(
\delta_{jn}\delta_{im}\left(N + 1 \right) - \delta_{ni}\delta_{jm}
- \delta_{nm}\delta_{ij} \right)$. The density matrix is obtained
as the expectation value of the density matrix in
(\ref{LiouvilleSolution}) according to
(\ref{AveragedMultiplIntFinal}):
\begin{equation}
\label{DensityMatrixAbstract} \rho\left(t \right) = \hat{P} \exp
\left(i \overline{\hat{X}} t - 1/2 \: \tau_{\rm corr} \hat{C} t
\right) \rho \left(0 \right)
\end{equation}
Note that the operator $\hat{X}$ is invariant under the
transformations from the orthogonal group $O\left(N \right)$, but
these transformations do not build irreducible representation, and
the second-order correlation function is not proportional to the
identity. However the expression (\ref{DensityMatrixAbstract}) can
be calculated in closed form and the final result is:
\begin{equation}
\label{DensityMatrixFinal}\rho_{ij}\left(t \right)  = \left(
\rho_{ij}\left(0 \right) - 1/N \:  \delta _{ij} \right) \exp
\left( i\left( E_{j} - E_{i} \right)t - \sigma ^{2}N\tau_{\rm
corr} t \right) + 1/N \: \delta _{ij}
\end{equation}
The density matrix of the unperturbed system is
$\rho_{ij}^{0}\left(t \right) = \rho_{ij}\left(0 \right) \exp
\left(i\left(E_{i} - E_{j}\right) t \right)$. The fidelity is $
f_{2}\left(t \right) = \rho_{ij}\left(t
\right)\rho_{ji}^{0}\left(t \right) = \left(1 - 1/N \right) \exp
\left(-\sigma^{2} N \tau_{\rm corr} t \right) + 1/N $. The
expectation value of the fidelity $f_{1}$ is obtained from
(\ref{AveragedMultiplIntFinal}) in a similar way using the
second-order correlation function of the hamiltonian
$\overline{H_{ik}H_{kj}} - \overline{H}_{ik}\overline{H}_{kj} =
\sigma^{2} \left(N + 1\right)\delta_{ij}$, and the result is $
\overline{f_{1}}\left(t \right) = \exp \left(-1/2 \:
\sigma^{2}\tau_{\rm corr}\left(N + 1 \right) t \right)$, which is
in a good agreement with the exact expression for $f_{2}\left(t
\right)$. Thus the fidelity decay rate which is obtained in exact
but somewhat complicated way and those which is obtained by simple
averaging differ only by the factor $N/(N + 1)$, which can be
explained by randomly changing phase of the fidelity $f_{1}$. The
asymptotic behaviour is different, but it is not considered here.
Further in this paper the expectation value of the fidelity
$f_{1}$ is used to estimate the fidelity decay rate.
\subsection*{3. Wilson loop}
\indent \indent Wilson loop is commonly used to test the confining
properties of Yang-Mills fields. Wilson loop is defined as the
trace of the multiplicative integral of gauge field $
\hat{A}_{\mu} \left( {\bf x} \right) $ over some closed contour
$\gamma$ in Minkowski space \cite{Sim1, Sim2}:
\begin{equation}
\label{WilsonLoop} W\left(\gamma \right) = 1/N_{c} \: {\rm Tr} \:
\hat{P} \exp \left( i g \int \limits_{\gamma} dx^{\mu}
\hat{A}_{\mu} \right)
\end{equation}
where the hat symbol denotes operators in the colour space,
$N_{c}$ is the number of colours and $g$ is the coupling constant.
If the ``area law'' holds for the Wilson loop $W\left( \gamma
\right)$, that is $W\left( \gamma \right) \sim \exp \left( -
\sigma S_{\gamma} \right)$ , $S_{\gamma} $ being the minimal area
of the surface spanned over the contour $\gamma$, quarks are said
to be tied by a string with the constant "tension'' $\sigma$
\cite{Sim1, Sim2}. If the gauge fields $ \hat{A}_{\mu} \left( {\bf
x} \right)$ form stochastic ensemble, the expectation value of the
Wilson loop is considered \cite{Sim1, Sim2}. In the case of
stochastic vacuum in QCD one usually chooses the curvature tensor
of gauge field $\hat {F}_{\mu \nu}  =
\partial _{\mu}  \hat {A}_{\nu}  -
\partial _{\nu}  \hat {A}_{\mu}  - ig\left[ {\hat {A}_{\mu} ,\hat
{A}_{\nu} }  \right] $ as a random variable \cite{Sim1,Sim2}. For
topologically trivial fields the multiplicative integral over the
contour $\gamma$ in (\ref{WilsonLoop}) can be represented as the
multiplicative integral over the surface spanned over the contour
$\gamma$ by applying the non-Abelian Stokes theorem \cite{Sim1,
Sim2, NAST}:
\begin{equation}
\label{eq3} W\left( \gamma \right) = 1/N_{c} \: {\rm Tr} \:
\hat{P} \exp \left( {i g \int \limits_{S_{\gamma}} dS^{\mu \nu}
\tilde {F}_{\mu \nu} } \right)
\end{equation}
where $\tilde {F}_{\mu \nu} \left( {\bf y} \right)  = \hat
{U}\left( {\bf x},{\bf y} \right)\hat {F}_{\mu \nu } \left( {\bf
y} \right)\hat {U}\left( {\bf y},{\bf x} \right)$ is the shifted
curvature tensor,\\ $\hat {U}\left( {\bf x},{\bf y} \right) = \hat
{P} \exp \left( i g \int \limits_{y}^{x} dx^{\mu} \hat {A}_{\mu}
\right)$ and ${\bf x}$ is an arbitrary point on the contour
$\gamma$. The dimensionality $d$ of the manifold $D$ in
(\ref{MultiplInt}) is $2$ in this case, and the condition of
gaussian dominance (\ref{GaussianDominance}) for the random field
$\hat {F}_{\mu \nu} \left({\bf x} \right) $ is:
\begin{equation}
\label{GaussDominanceQCD} g \sqrt {\overline {\hat {F}^{2}}} \cdot
l_{\rm corr}^{2} \ll 1
\end{equation}
\indent QCD is by definition invariant under local gauge
transformations. The transformation law for the curvature tensor
$\hat {F}_{\mu \nu}$ is $\hat {F}_{\mu \nu} \left({\bf x} \right)
\rightarrow \hat{T}^{\dag}({\bf x}) \hat {F}_{\mu \nu}\left({\bf
x} \right)\hat{T}({\bf x}) $, where $\hat{T}\left({\bf x} \right)
\in SU \left(3 \right)$. As the gauge symmetry should be also
preserved for the random field $\hat {F}_{\mu \nu} \left({\bf
x}\right)$, the probability measure on $\hat {F}_{\mu
\nu}\left({\bf x}\right)$ should be invariant under gauge
transformations. Now the results of section 1 can be applied to
calculate the expectation value of the Wilson loop. The random
field is the field of curvature tensor, the manifold $D$ is the
surface spanned over the contour $\gamma$, $dV$ is the element of
area, the invariance group $G$ is the group of local gauge
transformations, which obviously includes the subgroup of global
gauge transformations. The final expression for the expectation
value of the Wilson loop according to
(\ref{AveragedMultiplIntFinal}) is:
\begin{equation}
\label{eq9} \overline{W\left(\gamma \right)} = \exp \left( - 1/2
\: g^{2} \overline{\hat{F}^{2}} l_{\rm corr}^{2} S_{\gamma}
\right)
\end{equation}
Thus the Wilson area law for the Wilson loop holds, and the
gaussian-dominated stochastic vacuum possesses confining
properties. The string tension $\sigma$ is also obtained from
(\ref{eq9}): $\sigma = 1/2 \: g^{2} l_{\rm corr}^{2} F^{2}$.
\subsection*{4. Holonomic quantum computations}
\indent \indent Holonomic quantum computations are implemented as
unitary transformations of the vector space spanned on
eigenvectors corresponding to the degenerate energy level of the
hamiltonian $\hat{H}$ which depends on the set of external
parameters $\lambda^{\mu}$: $ \hat{H}\left(\lambda \right) =
\hat{U}\left(\lambda \right) \hat{H}_{0} \hat{U}^{\dag}
\left(\lambda \right)$, where $\hat{U}\left( \lambda \right)$ is
unitary. Typically degeneracy of the energy levels is the
consequence of existence of the invariance group $G$ of the
hamiltonian $\hat{H}\left( \lambda^{\mu} \right)$ \cite{HQC}.
Wigner's theorem states that if the hamiltonian of the system is
invariant under transformations $\hat{T}$ which build a
representation of the group $G$ on the Hilbert space of the
system, that is $\hat{T}^{-1}\hat{H}\hat{T}
 = \hat{H}$, each energy level is degenerate, and transformations
of the vector subspaces of the Hilbert space of the system which
correspond to the degenerate energy levels build irreducible
representations of the group $G$ \cite{Elliot}. Only one
degenerate energy level with eigenvectors which build a basis of
an irreducible representation is considered further. Holonomic
quantum computations are performed by changing the parameters
$\lambda^{\mu}$ adiabatically \cite{HQC}. In this case dynamical
effects can be neglected, and an arbitrary quantum gate is then
described by the multiplicative integral over some contour
$\gamma$ in the space of external parameters:
\begin{equation}
\label{HQCGATE} \hat{\Gamma}_{\gamma} = \hat{P} \exp \left( i \int
\limits_{\gamma} d\lambda^{\mu} \hat{A}_{\mu} \left(\lambda
\right)
 \right)
\end{equation}
where $A_{\mu; \: ij}\left(\lambda \right) = \left \langle
\phi_{i}\left(\lambda \right) \right |\hat{A}_{\mu}\left(\lambda
\right)\left | \phi_{j}\left(\lambda \right)
 \right \rangle  = - i \left \langle \phi_{i}\left(\lambda \right) \right |
\frac{\partial}{\partial \lambda^{\mu}} \left |
\phi_{j}\left(\lambda \right) \right \rangle$ is the adiabatic
connection on the space of external parameters and $ \left
|\phi_{i}\left(\lambda \right) \right \rangle$, $i = 1\ldots N$
are the eigenvectors which correspond to the degenerate energy
level.
\\ \indent It is interesting to note the close analogy
between the theory of gauge fields and the theory of holonomic
quantum computations. For example, in the theory of gauge fields
gauge transformations correspond to local transformations of basis
vectors in the colour space, and the transformation law for the
gauge field is $ \hat{A}_{\mu} \left({\bf x} \right) \rightarrow
\hat{T}^{-1}\left({\bf x} \right) \hat{A}_{\mu} \left({\bf x}
\right) \hat{T}\left({\bf x} \right) - i \hat{T}^{-1}\left({\bf x}
\right) \partial_{\mu} \hat{T}\left({\bf x} \right)$. Physical
observables obviously do not depend on the choice of basis
vectors, and the consequence is the gauge invariance of the
theory.  For holonomic quantum computations it is possible to
consider unitary transformations of basis vectors which correspond
to the degenerate energy level. These transformations in general
depend on the external parameters $\lambda^{\mu}$. The adiabatic
connection $\hat{A}_{\mu}\left(\lambda \right)$ transforms under
local transformations of basis vectors as $A_{\mu; \: ij}
\left(\lambda \right) \rightarrow - i \left \langle
\phi_{i}\left(\lambda \right) \right | \hat{T}^{-1}\left(\lambda
\right) \frac{\partial}{\partial
\lambda^{\mu}}\hat{T}\left(\lambda \right) \left |
\phi_{j}\left(\lambda \right) \right \rangle = \\
 - i \sum
\limits_{k,l = 1}^{N} \left \langle \phi_{i}\left(\lambda \right)
\right | \hat{T}^{-1}\left(\lambda \right) \left|
\phi_{k}\left(\lambda \right) \right \rangle \left \langle
\phi_{k}\left(\lambda \right) \right| \frac{\partial}{\partial
\lambda^{\mu}}\left| \phi_{l}\left(\lambda \right) \right \rangle
\left \langle \phi_{l}\left(\lambda \right)
\right|\hat{T}^{-1}\left(\lambda \right) \left |
\phi_{j}\left(\lambda \right) \right \rangle = \\
T^{-1}_{ik}\left(\lambda \right)A_{\mu; \: kl}\left(\lambda
\right)T_{lj}\left(\lambda \right) - i T^{-1} _{ik}\left(\lambda
\right)\frac{\partial}{\partial \lambda^{\mu}}T_{kj}\left(\lambda
\right)$, or, equivalently,\\ $ \hat{A}_{\mu} \left(\lambda
\right)\rightarrow \hat{T}^{-1}\left(\lambda
\right)\hat{A}_{\mu}\left(\lambda \right) \hat{T}\left(\lambda
\right) - i \hat{T}^{-1}\left(\lambda
\right)\frac{\partial}{\partial \lambda^{\mu}}\hat{T}\left(\lambda
\right)$ which is evidently the same as the transformation law for
the gauge field. The principle of gauge invariance can be
reformulated in this case as invariance of physical observables
with respect to the choice of irreducible representation of the
group $G$ among all equivalent irreducible representations.
Transitions between equivalent irreducible representations
correspond to unitary transformations of the basis vectors of the
representations \cite{Elliot}. \\ \indent The fidelity of
holonomic quantum computations with respect to errors in control
parameters is defined as the scalar product of the state vectors
which correspond to desired and actual operation. Using the
general expression for the quantum gate (\ref{HQCGATE}), the
fidelity of the holonomic quantum computation can be expressed in
terms of the multiplicative integral \cite{ Kuvshinov, Prosen}:
\begin{equation}
\label{HQCFidelity} f  = \left \langle \psi_{0} \right| \hat{P}
\exp \left( i \int \limits_{\delta \gamma} d \lambda^{\mu}
\hat{A}_{\mu} \right) \left| \psi_{0} \right \rangle
\end{equation}
where $\delta\gamma$ is the contour obtained by traveling forward
in the path $\gamma_{1}$ which corresponds to desired operation
and backward in the path $\gamma_{2}$ which corresponds to actual
operation and $ \left| \psi_{0} \right \rangle$ is the initial
state of the quantum register. As the fidelity is usually close to
unity (otherwise holonomic quantum computations will lead to
unpredictable results and will be therefore useless), the distance
between the contours $\gamma_{1}$ and $\gamma_{2}$ should be very
small. In this case it is possible to rewrite (\ref{HQCFidelity})
in the form where integration is performed over only one contour.
This is achieved by continuously deforming the contour
$\gamma_{2}$ to the contour $\gamma_{1}$ and simultaneously
changing the vector $\hat{A}_{\mu}$ on it in such a way that the
multiplicative integral in (\ref{HQCFidelity}) is not changed:
\begin{equation}
\label{ContinuousDeforming} \hat{P} \exp \left( i \int
\limits_{\gamma_{2}} d\lambda ^{\mu} \hat {A}_{\mu} \right) = \hat
{P} \exp \left( i \int \limits_{\gamma _{1}} d\lambda ^{\mu}
\left( \hat {A}_{\mu} + \delta \hat {A}_{\mu } \right)  \right)
\end{equation}
To obtain the correcting term $\delta \hat{A}_{\mu} \left(
\lambda^{\nu} \right)$ we multiply the
equation(\ref{ContinuousDeforming}) from the left by $ \left(
\hat{P} \exp \left( \int \limits_{\gamma_{1}} d \lambda^{\mu}
\hat{A}_{\mu} \right) \right)^{-1}$. The left side is transformed
by applying the non-Abelian Stokes theorem:
\begin{equation}
\label{ContinuousDeforming2}
\begin{array}{l}
\left( \hat{P} \exp \left( i \int \limits_{\gamma_{1}}d
\lambda^{\mu} \hat{A}_{\mu}\right) \right)^{-1}\hat{P} \exp \left(
i \int \limits_{\gamma_{2}} d\lambda ^{\mu} \hat {A}_{\mu} \right)
 = \\
\hat{P} \exp \left( i \int \limits_{\delta\gamma} d\lambda ^{\mu}
\hat {A}_{\mu}  \right) = \hat {P} \exp \left( i\int
\limits_{\gamma_{1}} d\lambda^{\mu} \tilde {F}_{\mu \nu}  \delta
\lambda^{\nu} \left( \lambda \right) \right)
\end{array}
\end{equation}
where $\delta \lambda^{\nu}$ is the distance between corresponding
points on the contours $\gamma_{1}$ and $\gamma_{2}$ and $\tilde
{F}_{\mu \nu}$ is the shifted curvature tensor defined as in
section 3 with the gauge field $\hat{A}_{\mu} \left({\bf x}
\right)$ replaced by the adiabatic connection $\hat{A}_{\mu}
\left(\lambda \right)$. The right side of
(\ref{ContinuousDeforming}) is transformed as:
\begin{equation}
\label{ContinuousDeforming3}
\begin{array}{l}
\left( \hat{P} \exp \left( i \int \limits_{\gamma_{1}} d
\lambda^{\mu} \hat{A}_{\mu}   \right) \right)^{-1} \hat{P} \exp
\left( i \int \limits_{\gamma_{1}}d\lambda ^{\mu} \left(\hat
{A}_{\mu} + \delta \hat {A}_{\mu} \right)  \right)=\\ \hat{P} \exp
\left( i \int \limits_{\gamma_{1}} d\lambda ^{\mu}
\hat{U}^{-1}\left(\lambda \right)\delta \hat{A}_{\mu}\hat{U}\left(\lambda
\right)    \right)
\end{array}
\end{equation}
where $\hat{U}\left(\lambda \right) = \hat{P} \exp \left( \int
\limits_{0}^{\lambda} d\lambda^{\mu} \hat{A}_{\mu} \right)$ is the
multiplicative integral of the adiabatic connection
$\hat{A}_{\mu}\left(\lambda \right)$ over the piece of the contour
$\gamma_{1}$ which is limited by the initial point and the point
$\lambda$. The correcting term $\delta\hat{A}_{\mu}$ in
(\ref{ContinuousDeforming}) is then obtained by comparing the
arguments of the multiplicative integrals in
(\ref{ContinuousDeforming3}) and (\ref{ContinuousDeforming2}):
\begin{equation}
\label{CorrectingTerm}
\delta\hat{A}_{\mu} \left( \lambda \right)
= \hat{F}_{\mu\nu} \left( \lambda \right) \delta\lambda^{\nu}
\end{equation}
The expression (\ref{HQCFidelity}) for the fidelity of holonomic
quantum computations is then rewritten as:
\begin{equation}
\label{FidelityHQC} f  = \left \langle \psi_{0} \right| \hat{P}
\exp \left( \int \limits_{\gamma_{1}} d\lambda^{\mu}
\hat{U}^{-1}\left(\lambda \right) \delta\hat{A}_{\mu}\left(
\lambda \right) \hat{U}\left(\lambda \right)  \right) \left|
\psi_{0} \right \rangle
\end{equation}
\indent The error $\delta\lambda$ is usually random. It is
therefore important to predict the fidelity decay rate from the
statistical properties of the errors. As in section 2, the
expectation value of the fidelity will be used to estimate the
fidelity decay rate. The expectation value of the multiplicative
integral in (\ref{FidelityHQC}) can be calculated using the
expression (\ref{AveragedMultiplIntFinal}). The manifold $D$ is
now the contour $ \gamma_{1}$, and $dV$ is the element of length
on it. One can expect that the probability measure on
$\delta\hat{A}_{\mu} \left(\lambda \right)$ does not depent on the
choice of the irreducible representation, hence the invariance
group is $G$, and the multiplicative integral in
(\ref{HQCFidelity}) according to section 1 is proportional to the
identity. Therefore the fidelity does not depend on the initial
state $\left| \psi_{0} \right \rangle$, and the final expression
for the expectation value of the fidelity is:
\begin{equation}
\label{HQCFidelityFinal}
\begin{array}{ll}
 f = \exp \left(-1/2 \: \lambda
_{\rm corr} \int \limits_{\gamma_{1}}
\overline{\delta\hat{A}_{\mu}\delta\hat{A}_{\nu}}n^{\mu}d\lambda^{\nu}
\right) =\\ \exp \left( -1/2 \: \lambda _{\rm corr} \int
\limits_{\gamma _{1}} d\lambda ^{\nu} \hat {F}_{\chi \alpha}
\left( {\lambda} \right)\hat {F}_{\nu \beta} \left( {\lambda}
\right) \overline {\delta\lambda^{\alpha}\delta\lambda^{\beta}}
n^{\chi} \right)
\end{array}
\end{equation}
where $n^{\chi}$ is the directing vector of the path $\gamma_{1}$
and $\lambda_{\rm corr}$ is the correlation length of the error
$\delta \lambda^{\nu} \left(\lambda \right)$ which indicates the
relative frequency of errors along the contour $\gamma_{1}$. The
expression (\ref{HQCFidelityFinal}) shows that the stability of
quantum computation essentially depends on:
\begin{itemize}
  \item the duration and the closeness of errors (the factor $\lambda_{\rm corr}$ in (\ref{HQCFidelityFinal}))
  \item the curvature tensor on the space of external parameters $\hat{F}_{\mu\nu}$
  \item the expectation value of the square of the magnitude of errors
  \item the length of the contour $\gamma_{1}$ in the space of external parameters
\end{itemize}
\subsection*{5. Fidelity in QCD}
\indent \indent Using the analogy between the theory of gauge
fields and the theory of holonomic quantum computations which was
pointed out in section 4 we define the fidelity of quark as the
scalar product of state vectors in the colour space $| q_{1}
\rangle $ and $| q_{2} \rangle $ which correspond to perturbed and
unperturbed motion:
\begin{equation}
\label{eq10}
 f = \langle q_{1}|q_{2} \rangle,
\quad \left| q_{1} \right\rangle, \left| q_{2} \right\rangle \in
C^{3}, \quad \langle q_{1}|q_{1} \rangle = \langle q_{2}|q_{2}
\rangle = 1
\end{equation}
\indent It is now possible to apply the results which were
obtained in previous sections and to obtain the fidelity decay
rates for different types of quark motion. First we consider the
motion of a coloured quark in different paths $\gamma _{1} $ and
$\gamma _{2} $ in the QCD stochastic vacuum. The paths start from
the point ${\bf x}$ and join in the point ${\bf y}$. In the
initial point ${\bf x}$ the state vector is $\left| q_{0}
\right\rangle $. In the limit of very massive quarks \cite{Sim1,
Sim2} evolution of the state vectors in the colour space is
described by the multiplicative integral introduced in
(\ref{WilsonLoop}):
\begin{equation}
\label{eq11} \left| {q_{k}}  \right\rangle = \hat {P} \exp \left(
i g \int\limits_{\gamma _{k}} dx^{\mu} \hat {A}_{\mu}
\right)\left| {q_{0}}  \right\rangle = \hat {U}\left( {\gamma
_{k}} \right)\left| {q_{0}}  \right\rangle, \quad k=1, 2
\end{equation}
The operators $\hat {U}\left( {\gamma _{1}}  \right)$ and $\hat
{U}\left( {\gamma _{2}}  \right)$ are unitary because $\hat
{A}_{\mu}  $ is hermitan. Taking this into account, one can
rewrite the expression for the fidelity as the multiplicative
integral over the closed contour $\gamma = \gamma_{1}
\bar\gamma_{2}$ obtained from the paths $\gamma _{1} $ and $\gamma
_{2} $ by travelling from the point ${\bf x}$ to the point ${\bf
y}$ in the path $\gamma_{1}$ and back to the point ${\bf x}$ in
the path $\gamma_{2}$:
\begin{equation}
\label{eq12} f = \left\langle q_{0}  \right|\overline {\hat
{U}\left( {\gamma _{1}} \right) \cdot \hat {U}^{ +} \left( {\gamma
_{2}}  \right)} \left| q_{0} \right\rangle = \left\langle q_{0}
\right|\overline {\hat {U}\left( \gamma \right)} \left| q_{0}
\right\rangle
\end{equation}
As in the section 2 we use the expectation value of the fidelity
to estimate the fidelity decay rate. All calculations which are
necessary have been already made in section 3. The multiplicative
integral in (\ref{eq12}) is proportional to the identity due to
the colour neutrality of the stochastic vacuum. The expression for
the fidelity then does not depend on the initial state $| q_{0}
\rangle$ because of the normalization. The condition of gaussian
dominance is (\ref{GaussDominanceQCD}). The final expression for
the fidelity of the quark moving in the gaussian-dominated
stochastic vacuum is:
\begin{equation}
\label{FidelityInQCDVacuum} f = \exp \left( - 1/2 \: g^{2} l_{\rm
corr}^{2} \overline{\hat{F}^{2}} S_{\gamma} \right )
\end{equation}
where $S_{\gamma}$ is the area of the surface spanned over the
contour $\gamma$. Thus for the gaussian-dominated stochastic
vacuum the fidelity for the quark moving in different paths decays
exponentially with the area of the surface spanned over the paths,
the decay rate being equal to the string ``tension'' $\sigma $.
This hints at the close connection between the stability of quark
motion and quark confinement.
\\ \indent Another possible situation,
which is more close to the standard treatment of the fidelity, is
realized when $\gamma _{1}$ and $\gamma _{2}$ are two random paths
in Minkowski space which are very close to each other. The
corresponding expression for the fidelity is similar to
(\ref{eq12}), but now the averaging is performed with respect to
all random paths which are close enough. All the necessary
calculations are similar to those in the section 4. The invariance
group is again the group of local gauge transformations. The final
expression for the fidelity in this case is:
\begin{equation}
\label{eq14} f = \exp \left(  - 1/2 \: g^{2}l_{\rm corr}
\int\limits_{\gamma _{1}} dx^{\nu} \tilde {F}_{\chi \alpha} \tilde
{F}_{\nu \beta}  v^{\chi}  \overline {\delta x^{\alpha} \delta
x^{\beta} } \right)
\end{equation} where $\delta
x^{\alpha} $ is the deviation of the path $\gamma _{2} $ from the
path $\gamma _{1} $, $v^{\chi}$ is the four-dimensional velocity
and  $l_{\rm corr}$ is the correlation length of perturbation of
the quark path expressed in units of world line length. For
example, if unperturbed path is parallel to the time axis in
Minkowski space, the quark moves randomly around some point in
three-dimensional space. The fidelity in this case decays
exponentially with time, as would be expected.
\subsection*{Conclusions}
\indent \indent In this paper the interactions between different
quantum systems and the stochastic environment was studied. It was
shown that the approach described in section 1 is universal and
can be applied to a wide class of systems in different branches of
physics - from simple quantum systems to quantum computer and
gauge field theories. It is shown that existence of the invariance
group of the environment to a large extended determines the
behavior of quantum system. This phenomenon is intensively
investigated in the Random Matrix Theory \cite{RMT}, where the
statistical properties of eigenvalues of operators are considered.
The stochastic ensemble of fields which is invariant under the
action of some group can be regarded as the straightforward
generalization of orthogonal or unitary ensembles (GOE, GUE) in
the Random Matrix Theory. However it was sufficient to require
invariance under the global transformations only to obtain such an
important properties of stochastic ensemble as proportionality of
the correlators to the identity. The consequences of invariance
under local gauge transformations, as well as the role of gauge
invariance in the emergence of chaotic behaviour require further
investigations. It was also shown that the fidelity decay rate for
quarks moving in QCD stochastic vacuum is equal to the "string
tension" which characterizes the confining properties of
Yang-Mills fields. Therefore a relation exists between the
property of quark confinement in stochastic vacuum and the
stability of quark motion. This fact can be related to possible
mechanisms of quark confinement.

\end{document}